\documentclass[amsmath,twocolumn,showpacs,aps,prl]{revtex4}
\usepackage{bm}
\usepackage{graphics}
\usepackage{graphicx}

\begin{document}

\title{Aharonov-Bohm effect in superconducting  LOFF state}

\author{A. A. Zyuzin} \email{alexanderzyuzin@gmail.com} \author{A. Yu. Zyuzin}

\affiliation{ A. F. Ioffe
Physico-Technical Institute of Russian Academy of Sciences, 194021
St. Petersburg, Russia}


\begin{abstract}
We study AB oscillations of transition temperature,
paraconductivity and specific heat of thin ring in the regime of
inhomogeneous Larkin - Ovchinnikov - Fulde - Ferrell
superconducting state. We found that in contrast to uniform
superconductivity magnetic flux might increase the critical
temperature of LOFF state. Degeneracy of the inhomogeneous
superconducting state reveals in double peak structure of AB
oscillations.

\end{abstract}
\maketitle

A magnetic field destroys superconductivity either by orbital or
paramagnetic pair-breaking effects. The Chandrasekhar - Clogston
pair breaking limit takes place when paramagnetic energy coincides
with the superconducting condensation energy. Larkin and
Ovchinnikov \cite{bib: LO}, Fulde and Ferrel \cite{bib: FF}
predicted the existence of the nonuniform superconducting state in
ferromagnetic superconductors at low temperatures above the
paramagnetic limit (see for a review \cite{bib: LOFF1, bib:
LOFF2}). This so-called LOFF state is a result of Cooper pairing
with nonzero momentum and has a lower energy compared to the
uniform superconducting state.

Appearance of the LOFF state is related to change in the sign of
coefficient $\beta$ at the gradient term of the Ginzburg - Landau
(GL) free energy functional $\beta |\mathbf{\nabla}\Psi|^2$, where
$\Psi$ is the order parameter. Coefficient $\beta$ being a
function of Zeeman energy $\mu_B H$,  becomes negative at low
temperatures $T < 0.56 T_c(0)$ and high magnetic fields $H>1.07
T_c(0)/\mu_B$, where $T_c(0)$ is the critical temperature at zero
magnetic field, signalling of the formation of nonuniform LOFF
state. As a result one has to take into account higher terms in
the GL functional expansion $ |\mathbf{\nabla}^2\Psi|^2$. This
effect is very sensitive to impurities \cite{bib: Aslamazov} and
moreover usually orbital pair breaking effect dominates over the
paramagnetic limit.

Despite the LOFF state has been theoretically predicted almost
40ty years ago, only recently LOFF phase was found in heavy-
fermion compound $\mathrm{CeCoIn_5}$ and organic superconductors
like $\mathrm{\lambda-(BETS)_2 FeCl_4}$. The experimental evidence
of the LOFF state based on the specific heat measurements
\cite{bib: HF1, bib: HF3, bib: HF4} and nuclear magnetic resonance
\cite{bib: HF2} were presented for heavy- fermion superconductor.
The signature of phase transition between LOFF state and the
homogenous superconducting state was reported for organic
superconductors \cite{bib: OS1, bib: OS2, bib: OS3, bib: OS4}.
These experiments were focused on the identification of the phase
transition inferred from a kink of thermal conductivity \cite{bib:
OS3}, observation of peculiar properties -dip structures- in the
resistance \cite{bib: OS2} and changes in the rigidity of the
vortex system \cite{bib: OS4}. The thermodynamic evidence of the
existence of narrow intermediate state (attributed to LOFF) which
separates the uniform superconducting state and normal state based
on specific heat measurements was presented in paper \cite{bib:
OS1}.

In the paper \cite{bib: LOFF_Fluct} crossovers between different
fluctuational regimes of paraconductivity and specific heat in the
vicinity of the LOFF transition were discussed. Authors showed
that these fluctuational contributions have specific temperature
dependencies compared to the case of uniform superconductivity and
could serve as an additional indicator of the LOFF state.

\begin{figure}[t] \centering
\includegraphics[width=8cm]{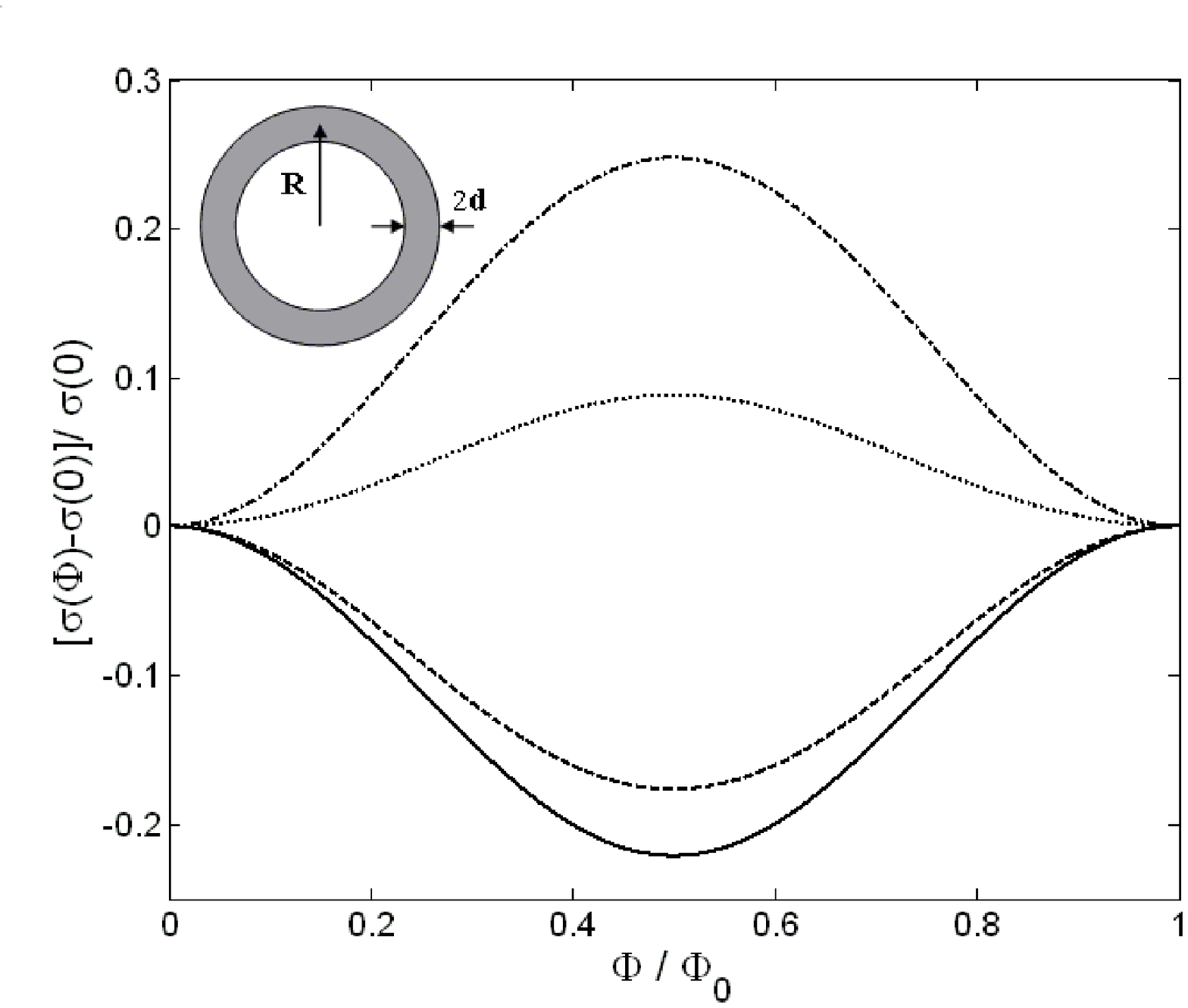} \caption{Magnetic flux dependence
of paraconductivity in the vicinity of LOFF transition
$z=[\beta^2/(4\alpha \delta)]^{1/4} = 20 \gg 1$ for different
radiuses of the ring $R$. Here solid line, dashed line, dash-dot
line, dotted line corresponds to $q = 2 \pi R
\left(\alpha/2|\beta|\right)^{1/2} = [5; 5.1; 5.56; 5.57]$
respectively} \label{fig:1}
\end{figure}

In the present letter we consider the Aharonov- Bohm effect in
superconducting ring of radius $R$, (see inset in
fig.\ref{fig:1}). We examine the superconducting fluctuations in
this system in the vicinity of the LOFF transition. It will be
shown that the transition temperature is an oscillating function
of the radius of the ring and can be increased by applied magnetic
flux. The detailed analysis of the paraconductivity and specific
heat for the superconducting ring will be given.

Because of the Little-Parks effect \cite{bib: LP1, bib: LP2} the
superconducting transition temperature oscillates with the applied
magnetic flux $\Phi/\Phi_0$ through the ring, where $\Phi_0 =
\pi/e$ is the flux quantum. We show that for the case of
metal-LOFF transition when parameter $\beta=-|\beta|$ becomes
negative, critical temperature is given as
\begin{eqnarray}\label{Tc}\nonumber
T^{*}_c(\Phi)=T_c (H)+\frac{\beta^2}{4a\delta}
-\frac{\delta}{aR^4}\left(\frac{\Phi}{\Phi_0}\right)^2 -
\\
-\frac{\delta}{aR^4}\textrm{min}\left( [n-\frac{\Phi}{\Phi_0}]^2
-\frac{R^2|\beta|}{2\delta}\right)^2
\end{eqnarray}
Where $\delta$ is a coefficient at the term $|\mathbf{\nabla}^2
\Psi|^2$ of GL functional and where integers $n$ are defined in
order to satisfy the minimum of the free energy: $E_n=a(T-T^{*}_c
(\Phi) )$. Depending on the ratio $\Phi/\Phi_0$ appropriate $n$
which satisfy expression \ref{Tc} jump between the integer parts
of two values
\begin{equation}\label{choice}
\pm\sqrt{\frac{R^2|\beta|}{2\delta}}+\frac{\Phi}{\Phi_0}
\end{equation}
This leads to oscillations of transition temperature with magnetic
flux or with ring's radius $R$.

\begin{figure}[h] \centering
\includegraphics[width=7cm]{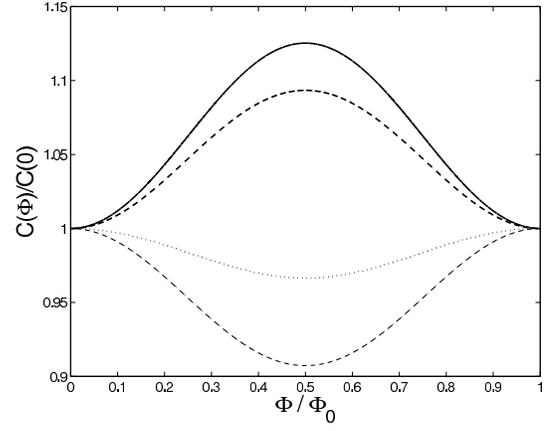} \caption{Magnetic flux dependence
of specific heat in the vicinity of LOFF transition $z = 20$ for
different radiuses of the ring $R$. Here solid line, dashed line,
dash-dot line, dotted line corresponds to $q = [5; 5.1; 5.6;
5.57]$ respectively} \label{fig:2}
\end{figure}

\begin{figure}[t] \centering
\includegraphics[width=7.5cm]{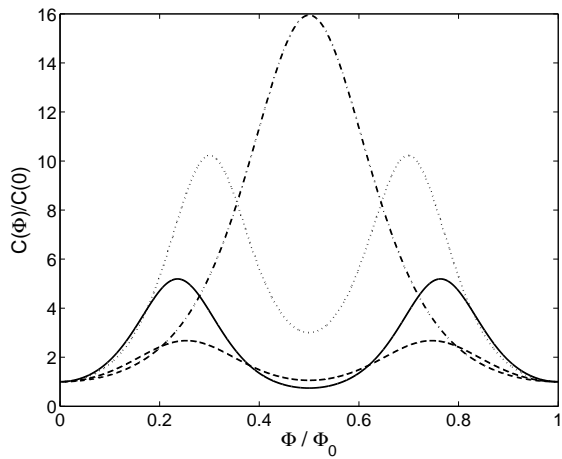} \caption{Magnetic flux dependence
of specific heat in the vicinity of LOFF transition for
$|T-T_c|\ll \beta^2/a\delta$, $(z = 20)$ for different radiuses of
the ring $R$. Here solid line, dashed line, dash-dot line, dotted
line corresponds to $q= 2 \pi R
\left(\alpha/2|\beta|\right)^{1/2}= [0.996; 1.4; 1.145; 0.995]$
respectively} \label{fig:3}
\end{figure}

\begin{figure}[h] \centering
\includegraphics[width=7.5cm]{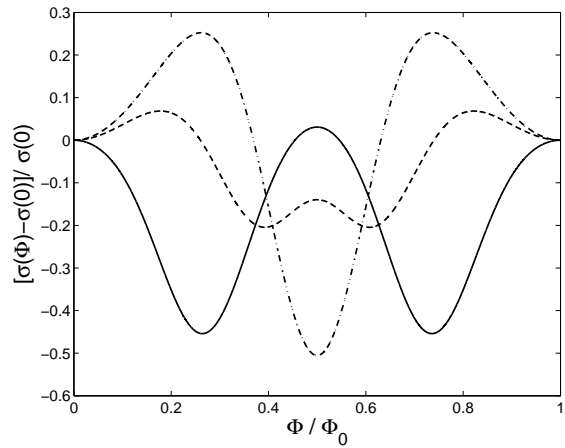} \caption{Magnetic flux dependence
of paraconductivity in the vicinity of LOFF transition $z = 20$
for different radiuses of the ring $R$. Here solid line, dashed
line, dash-dot line corresponds to $q = [1.875; 1.97; 1.975]$
respectively} \label{fig:4}
\end{figure}

It is seen that oscillations near LOFF behave in qualitatively
different way than in the vicinity of uniform superconductor
transition. Indeed, firstly, the critical temperature
$T^{*}_c(\Phi)$ is an oscillating function of the radius of the
ring $R$. Secondly, an applied magnetic flux in the regime of LOFF
transition can increase the critical temperature depending on the
sample size or on the free energy expansion parameters. These
effects are shown in fig.\ref{fig:1} and fig.\ref{fig:2} where the
magnetic flux dependencies of paraconductivity and specific heat
are presented for a set of rings with different radiuses. From eq.
\ref{choice} it is also seen that magnetic flux shifts the
degeneracy of superconducting LOFF state which might result in two
peaks in magnetic flux dependency of paraconductivity and specific
heat (see fig.\ref{fig:3} and fig.\ref{fig:4}).
Now let us tern to the detailed calculations. Consider free energy
density of a thin superconducting ring such that order parameter
is constant over the cross section. The spatial dependence of the
order parameter over the cross section of the ring will be
discussed at the end of the letter. Above the superconducting
transition one could use quadratic order parameter approximation
\begin{equation}
F=\Psi^* \{\tilde{\alpha} + \beta \mathbf{D}^2 + \delta
[(\mathbf{D}^2)^2 + (\Phi/(R^2 \Phi_0))^2 ]\}\Psi
\end{equation}
where $\mathbf{D}=-i\mathbf{\nabla} - 2e\mathbf{A}$, while tangent
component of the vector potential is given as
$A_{\varphi}=\Phi/(R\Phi_0)$, $\Psi$ is complex order parameter,
$\Phi$ is the flux of the magnetic field through the ring.
Coefficients $\tilde{\alpha} = a(T-T_c (H))$, $\beta$ and $\delta$
depend on the exchange magnetic field and temperature (see for
example \cite{bib: Kachkachi}).
Expanding order parameter in terms of Fourier series so that
\begin{equation}
\Psi = \sum_n \Psi_n e^{i\varphi n}
\end{equation}
we find for the free energy density
\begin{equation}
F= \sum_n E_n |\Psi_n|^2
\end{equation}
where spectrum of the fluctuations is given as
\begin{equation}
E_n = \alpha +\frac{\delta}{R^4}\left( [n-\frac{\Phi}{\Phi_0}]^2
+\frac{R^2\beta}{2\delta}\right)^2
\end{equation}
and $\alpha=\tilde{\alpha} - \frac{\beta^2}{4\delta}
+\frac{\delta}{R^4}\left(\frac{\Phi}{\Phi_0}\right)^2$. For ring
geometry the expressions for the paraconductivity and the specific
heat are \cite{bib: Larkin_Varlamov}
\begin{equation} \label{sigma}
\sigma = \frac{\pi e^2 a T_c}{R}\sum_{n} \frac{[\frac{\beta}{R^2}
(n-\frac{\Phi}{\Phi_0})
+\frac{2\delta}{R^4}(n-\frac{\Phi}{\Phi_0})^3]^2} {E_{n}^3}
\end{equation}
and
\begin{equation}\label{s_heat} C = \frac{a^2 T^{2}_{c}}{R}\sum_{n}
\frac{1} {E_{n}^2}
\end{equation}
Let us consider the case of metal-LOFF transition which
corresponds to negative $\beta=-|\beta|$. Performing Poisson
summation we obtain the expression for the paraconductivity
\begin{equation}\label{int_g}
\sigma=2\frac{(2\pi e)^2}{Rp}\frac{aT_c}{\alpha} \sum_k
\int_{-\infty}^{\infty} dt \frac{t^2(t^2 - z^2)^2
e^{ik(2\pi\Phi/\Phi_0 +tp)}}{[1+(t^2-z^2)^2]^3}
\end{equation}
and specific heat
\begin{equation}\label{int_c}
C= \frac{p}{2\pi R}\frac{(aT_c)^2}{\alpha^2}
\sum_k\int_{-\infty}^{\infty}dt \frac{ e^{ik(2\pi\Phi/\Phi_0
+tp)}}{[1+(t^2-z^2)^2]^2}
\end{equation}
where $z=[\beta^2/(4\alpha \delta)]^{1/4}$ and $p= 2\pi R
[\alpha/\delta]^{1/4}$.

First we will examine the limit $\beta^2\gg \alpha \delta$ which
corresponds to temperatures close to transition $|T-T_c| \ll
\beta^2/a\delta$. We also suggest the superconducting ring being
relatively large
\begin{equation}
R\gg\left(|\beta|/ \alpha\right)^{1/2}
\end{equation}
This condition allows one to keep modes with $k=0$ and $k=1$ in
equations (\ref{int_g}) and (\ref{int_c}) since higher modes will
be exponentially freezed. In this regime we obtain equation for
the paraconductivity that is given as
\begin{eqnarray}\label{sigma_minus} \nonumber
\sigma \simeq \frac{e^2}{R} \left(
\frac{|\beta|}{R^2aT_c}\right)^{1/2} |1-T/T_c|^{-3/2}\times
\\ \times \left[1- 2q^2
\cos{\left(2\pi\Phi/\Phi_0\right)}\cos{(\phi)}e^{-q}\right]
\end{eqnarray}
while the specific heat is given by the following expression
\begin{eqnarray}\label{sc_minus}\nonumber
C \simeq \frac{1}{R} \left( \frac{R^2aT_c}{|\beta|}\right)^{1/2}
|1-T/T_c|^{-3/2}\times
\\ \times \left[1+ 2q\cos{\left(2\pi\Phi/\Phi_0\right)}\cos{(\phi)}e^{-q}\right]
\end{eqnarray}
where we introduced parameters $\phi=pz= 2\pi R
\left(\frac{|\beta|}{2\delta}\right)^{1/2} $ and $q =p/2z= 2 \pi R
\left(\frac{\alpha}{2|\beta|}\right)^{1/2}$.
Depending on the sign of the $\cos{(\phi)}$ applied magnetic flux
can either increase or decrease the conductivity or specific heat.
That is in contrast to the case of normal to uniform
superconductor transition in superconducting rings where for
example specific heat always decreases with applied magnetic flux.
The different sign in expressions (\ref{sigma_minus}) and
(\ref{sc_minus}) reflects the different origins of specific heat
which is associated with the critical temperature behavior and
transverse paraconductivity which is a spectral structure
dependent quantity \cite{bib: Glazman_Zyuzin, bib:
Buzdin_Varlamov}.

Another interesting feature of the LOFF phase arises for the
intermediate values of the radius of the ring when $R\sim
\left(|\beta|/ \alpha \right)^{1/2}$. This situation is
illustrated in fig.\ref{fig:3} and fig.\ref{fig:4} where the
magnetic flux dependencies of the specific heat and
paraconductivity are presented for different radiuses of the ring.
One sees the crossover between different regimes of the
oscillations behavior, in particular, from the one-peak per period
into two peaks. Magnetic flux doubles the peaks by removing the
degeneracy of the superconducting state eq. (\ref{choice}). This
effect becomes more pronounced in the case of very small rings
\begin{equation}
R\ll\left(|\beta|/\alpha\right)^{1/2}
\end{equation}
where one will observe strong fluctuations of both
paraconductivity and specific heat. The superconducting ring
effectively becomes a quasi-zero dimensional system and one yields
the following expression for specific heat
\begin{equation}
C\simeq \frac{1}{R} |1-T/T_c|^{-2} \left[f(\phi +\frac{2\pi
\Phi}{\Phi_0})+f(\phi -\frac{2\pi \Phi}{\Phi_0}) \right]
\end{equation}
where
\begin{equation}
f(\phi +\frac{2\pi \Phi}{\Phi_0}) \simeq \frac{q^4}{4[1+q^2/2 -
\cos{(\phi + 2\pi\Phi/\Phi_0)}]^2}
\end{equation}
Note, that phase shifts $\phi \pm 2\pi \Phi/\Phi_0$ are equal to
the values of the phase in equation (\ref{choice}). Magnetic flux
removes the degeneracy of the superconducting state which reveals
in two-peak oscillations.

Now let us consider high temperatures regime far away from
transition where $|T-T_c| \gg \beta^2/a\delta$ or equivalently the
regime where coefficient $\beta\rightarrow 0$ vanishes. And let
the radius of the ring be $ R\gg \left(\delta/\alpha\right)^{1/4}
\gg \left(|\beta|/\alpha\right)^{1/2}$. We obtain expressions for
paraconductivity
\begin{eqnarray}\nonumber
\sigma \simeq \frac{e^2}{R} \left(\frac{\delta}{aT_c
R^4}\right)^{1/4}
|1-T/T_c|^{-5/4}\times \\
\times [1-\sqrt{2}p^2 e^{-p/\sqrt{2}} \sin{(p/\sqrt{2} +\pi/4)}
\cos{(2\pi\Phi/\Phi_0)}]
\end{eqnarray}
and specific heat
\begin{eqnarray}\nonumber
C \simeq \frac{1}{R}
\left(\frac{aT_cR^4}{\delta}\right)^{1/4}|1-T/T_c|^{-7/4} \times
\\ \times[1+\frac{2\sqrt{2}p}{3}e^{-p/\sqrt{2}}\sin{(p/\sqrt{2})}\cos{(2\pi\Phi/\Phi_0)}]
\end{eqnarray}
\begin{figure}[t] \centering
\includegraphics[width=7.5cm]
{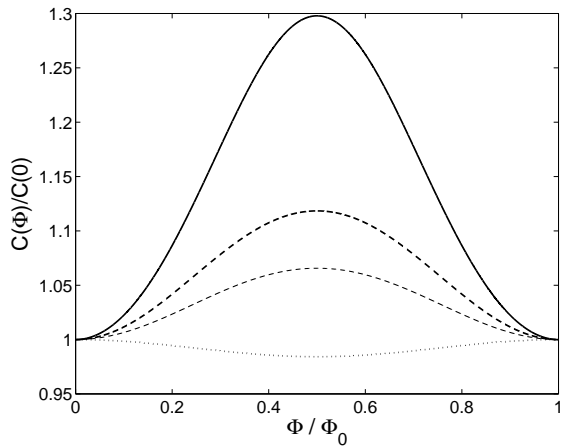} \caption{Magnetic flux dependence of specific heat in
the regime $|T-T_c|\gg \beta^2/a\delta$ for $z = 0.05$. Here solid
line, dashed line, dash-dot line, dotted line corresponds to $q =
[5; 7; 7.5; 9.5]$ respectively} \label{fig:5}
\end{figure}
Again, one sees (fig.\ref{fig:5}) that these fluctuational
contributions in the high temperatures regime also depend on the
random phase. Moreover, different temperature dependencies is an
additional property of the LOFF state \cite{bib: LOFF_Fluct} under
these conditions.

Now, we discuss possible effects coming from spatial dependence of
the order parameter over the cross section of the ring $2d$.
Complex order parameter written in cylindrical coordinates takes
the form $ \Psi(\phi ,\rho) = e^{i n\phi} f(\rho)$ where $f(\rho)$
is the Landau wave function. Then again the critical temperature
should be found by taking the minimum with respect to $n$ of the
free energy and we obtain
\begin{equation}
E_n = \alpha +\frac{\delta}{R^4} \min{[(n-\Phi/\Phi_0)^2 + \beta
R^2/2\delta + g]^2}
\end{equation}
where $g$ is a function of the ring thickness
\begin{equation}
g=\frac{R^2}{d^2}\left(\frac{\varphi}{\Phi_0}\right)^2+\frac{d^2n^2}{3R^2}
\end{equation}
and where $\varphi$ is the magnetic flux over the cross section of
the ring. The critical temperature below which LOFF modulation
appears is defined by the condition
\begin{equation}
\beta+ \frac{2 \delta }{3
d^2}\left(\frac{\varphi}{\Phi_0}\right)^2 = 0
\end{equation}
From this equation one sees that due to the orbital effect  LOFF
critical temperature decreases since now it is not enough for $\beta$ to
change sign but $\beta < - \frac{2\delta}{3
d^2}\left(\frac{\varphi}{\Phi_0}\right)^2$.

Finally, we focus on the validity of the Gaussian approximation
used in AB effect. Indeed, Brazovskii \cite{bib: Brazovskii}
showed that the critical fluctuations could be essential in LOFF
like systems and could lead to the first-order type transition.
The width of this critical fluctuations region (given by the
Levanyuk- Ginzburg parameter) increases \cite{bib: LOFF_Fluct}
compared to the uniform superconductor- metal transition. However,
our main result- the increase of the superconducting transition
temperature and double peaks in oscillations- is the effect of the
magnetic field that removes the degeneracy of the superconducting
state above transition. We note, that critical fluctuations do not
change the way of removing this degeneracy.

In conclusion, we have shown that an applied magnetic flux through
the thin ring in the vicinity of the LOFF transition can increase
the critical temperature. We calculated expressions for the
paraconductivity and specific heat in this regime. Both values
exhibits double peak oscillations in contrast to usual Little -
Parks effect in the normal to uniform superconducting transition.

We thank P. Fulde and V.I. Kozub for helpful discussions and
valuable questions. This work was financially supported by Dynasty
foundation, INTAS Grant 05- 109- 4829 and RFFI Grant 06- 02-
17047.

\end{document}